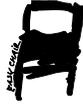

EasyChair Preprint
№ XXX

# Evaluation of Preprocessing Techniques for U-Net Based Automated Liver Segmentation

Muhammad Islam, Kaleem Nawaz Khan and Muhammad Salman Khan

March 26, 2021

# Evaluation of Preprocessing Techniques for U-Net Based Automated Liver Segmentation

Muhammad Islam
*Department of Electrical Engineering,*
*University of Engineering and Technology,*
Peshawar, Pakistan
m.islam5588@gmail.com

Kaleem Nawaz Khan
*Artificial Intelligence in Healthcare,*
*IIPL, NCAI, UET*
Peshawar, Pakistan
kaleemnawaz@uetmardan.edu.pk

Muhammad Salman Khan
*Department of Electrical Engineering,*
*Artificial Intelligence in Healthare, IIPL,*
*National Center of Artificial Intellgience,*
*Univerity of Engineering and Technology,*
Jalozai campus, Peshawar, Pakistan
salmankhan@uetpeshawar.edu.pk

*Abstract*—To extract liver from medical images is a challenging task due to similar intensity values of liver with adjacent organs, various contrast levels, various noise associated with medical images and irregular shape of liver. To address these issues, it is important to preprocess the medical images, i.e., computerized tomography (CT) and magnetic resonance imaging (MRI) data prior to liver analysis and quantification. This paper investigates the impact of permutation of various preprocessing techniques for CT images, on the automated liver segmentation using deep learning, i.e., U-Net architecture. The study focuses on Hounsfield Unit (HU) windowing, contrast limited adaptive histogram equalization (CLAHE), z-score normalization, median filtering and Block-Matching and 3D (BM3D) filtering. The segmented results show that combination of three techniques; HU-windowing, median filtering and z-score normalization achieve optimal performance with Dice coefficient of 96.93%, 90.77% and 90.84% for training, validation and testing respectively.

I. INTRODUCTION

Liver is a vital and the second largest organ in human body [1], therefore liver cancer and related diseases need to be diagnosed earlier that can assist to prevent complete liver failure, a cause of death. For clinical practice such as liver disease diagnosis, treatment planning, follow-up of patients and surgery, the commonly used modalities as medical imaging are Computed Tomography (CT) and Magnetic Resonance Imaging (MRI). The abnormalities in liver can be investigated by using manual segmentation of liver in CT image. However, due to large amount of CT volumetric data and other constraints, it is impossible for radiologists to analyze the data in short space of time. To assist radiologists, many automated liver segmentation algorithms have been developed using machine learning methods.

Recently, deep learning has improved the segmentation task and achieved great performance in medical image analysis [2]. Deep learning has been enabled the computers to learn the features in medical images that represent the problem [3]. Applying different deep learning techniques, the main tasks: segmentation, classification and detection have been performed with improved results [4]. Segmentation of medical image assists in medical finding and quantitative analysis of an organ of interest. Many researchers have been focused to automate the segmentation task using deep learning techniques. One of the deep learning models is U-Net architecture [5], shown state-of-the-art performance in medical image segmentation. However, due to certain limitations, automated segmentation of liver from CT images is still a challenging task. Some of these constraints are similarity of pixel values of liver with other adjacent abdomen organs, low contrast, intensity inhomogeneity and related noises.

Preprocessing is a preliminary step in deep learning to prepare smooth data for later stages. It plays a vital role in medical image segmentation, in which the acquired raw medical images are processed to assist the learning model in distinguishing the target organ from other body parts. In CT image, there are several organs having similar intensity value with liver, which make differentiation a difficult task for learning model. In this regard, HU windowing is considered the most important preprocessing technique and the researchers who study liver in medical images using deep learning, often limit the HU values range to focus on the target organ [6-7]. The quality of medical images obtained by different modalities can also be degraded by poor contrast, which can be improved by applying the adaptive histogram equalization technique, results in better segmentation at later stage [8]. Similarly, CT images are also subjected to various noises up to some extent, which degrade the quality of images. Hence, different denoising techniques are proposed in literature to obtain denoised CT images [9-10]. Another preprocessing step suggested in literature is data normalization, often employed to changes the range of pixel intensity values to a common scale. In summary, preprocessing techniques are essential and widely used to prepare CT data, which afterward fed to a learning model for automated liver segmentation.

The paper is organized in five different sections. After introduction to the problem in section I, section II presents the detailed literature review and related work. Proposed methods are discussed in section III. The results produced by the proposed approach are discussed and analyzed, in section IV. Finally, the last section concludes the paper and provides future directions.

## II. RELATED WORK

Segmentation of liver and liver tumor is a challenging task. Many automated and interactive methods have been introduced for fast and accurate extraction of liver and liver lesions from CT volumes. Some of the methods developed so far are texture-based methods [11], and statistical shape models presented at a Grand Challenge benchmark [12]. Furthermore, the graph-cut and machine learning methods have also been employed to segment liver and liver tumor [13-15]. However, these methods are slow and not so robust on real-life CT data having low contrast and heterogeneous, therefore, not widely used in clinical practice.

In the last few years, deep convolutional neural networks (CNN) have become a main focus for research community due to its capability to solve computer vision tasks. CNN is typically designed using three main layers that constitutes convolutional, pooling and fully connected neural network layers [16]. Long et al. [17], proposed the fully convolutional network (FCN), an advanced version of CNN. The tasks of semantic segmentation in medical imaging have been also successfully done using similar network architectures [18-20]. One of the networks based on the principal of fully convolutional networks known as U-Net [5]. It consists of contracting path and expansive path for features extraction and image reconstruction respectively.

In segmentation task, U-Net performs accurate localization by combining low- and high-level features in image, which has been made possible by addition of skip connection. Due to outstanding performance, U-Net has been proposed by many researchers as a learning model for automated segmentation of liver in CT images. In [21], Christ et al. implemented the U-Net architecture for liver and liver tumor segmentation, followed some preprocessing techniques on CT dataset. They preferred the cascaded approach of two cascaded U-Nets in two steps. In step 1, it was ensured to detect and segment only liver from abdominal CT scan by applying U-Net model to learn. In step 2, U-Net was trained that learned only lesion separation from extracted liver in step 1. In [22], the authors perform the liver segmentation task using 2D U-Net. Although, 3D based deep learning perform better, but due to computation limitation, 2D based methods are preferred to keep training fast.

The CT images acquired from different clinical sites with different protocols and various CT scanners often have different contrast level and have associated artifacts or noises. These related noises may degrade the performance of deep learning models while segmenting liver and liver tumor from CT images [23]. In order to refine CT images, several preprocessing techniques have been used in literature. HU [24] value is often applied and have been proposed by most of the researchers for segmentation of liver and liver lesion from medical images. To analyze the CT image, the range of HU value is limited to make the specific organ more visible and focus the target region. Some of the HU value ranges observed in literature were [-100, 400] in [25], a range of [-75, 175] in [26] and a range of [-200, 250] in [27].

Image contrast is an important visual feature that decides the quality of image. In computed tomography (CT), the image may have low contrast, which is a dominant artifact that degrades the image quality and cause a hinder in extracting image information [28]. The problem has been addressed by using histogram-based methods with combination of HU windowing [6, 29]. The histogram-based techniques are adaptive histogram equalization (AHE) and its variant contrast limited adaptive histogram equalization CLAHE [17]. Another preprocessing technique is data normalization, often applied in deep learning networks to change range of values to a common scale. Commonly used techniques for CT images are z-score and min-max normalization [30]. For medical image semantic segmentation, all subjects can be brought to similar distribution using z-score normalization with computing mean and standard deviation [31].

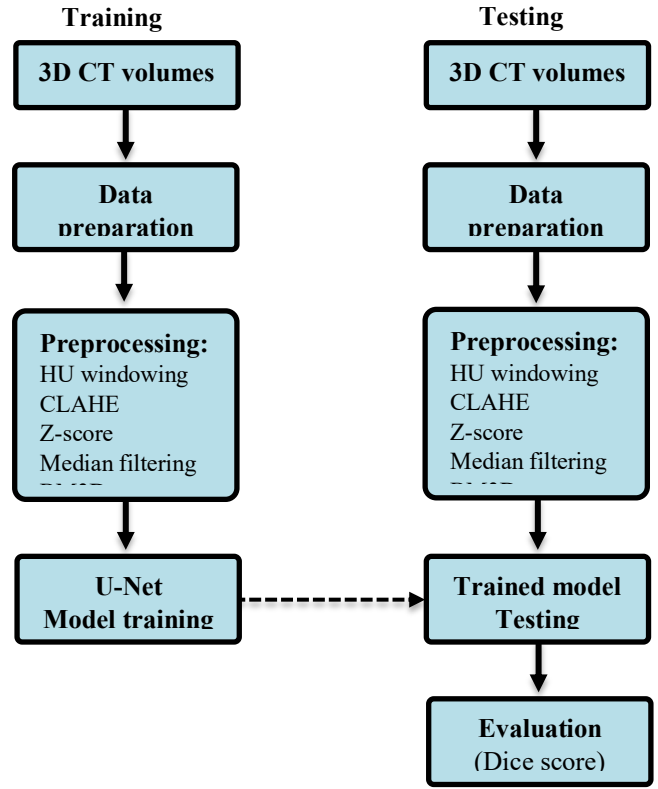

**Fig. 1.** Flow diagram of the proposed methodology

Medical images such as computed tomography (CT) play a vital role in clinical investigation. However, quality of these images is degraded by noises, causes a hindrance in abdomen and tumor segmentation. The noises can be eliminated by passing the CT images to denoising techniques such as Median filtering and Block-Matching and 3D filtering (BM3D). BM3D increases signal-to-noise ratio and enhanced spatial resolution, which improved training accuracy at later stage [32]. Median filter suppresses the noise while retaining the edge information within image and has advantage over other denoising techniques such as Mean filters and Gaussian filters which blur the edges of image [33].

## III. Proposed Methodology

The proposed methodology consists of two major phases, i.e., one is to apply various preprocessing techniques separately and jointly to enhance CT images. Second step is to use the preprocessed dataset with a deep learning algorithm to evaluate the performance of optimal preprocessing techniques for liver segmentation. This section first describes and discusses the dataset used for this research. After that various preprocessing techniques applied and different parameters used are discussed. In last the deep learning paradigm U-Net architecture is discussed and its performance is evaluated with all the preprocessing techniques under investigation. The proposed framework is depicted in Fig. 1 in which the prepared dataset is preprocessed by various preprocessing techniques separately and jointly. In training phase, deep learning model was trained using preprocessed data and finally the trained model is passed on to testing phase for performance evaluation.

*A. Dataset*

The proposed method is tested and its performance is evaluated by extensive experiments on the publicly available competitive dataset of MICCAI 2017 Liver Tumor Segmentation (LiTS) Challenge [36]. Each volume has different number of axial slices with resolution size of 512x512 pixel and varying thicknesses of slices. The experiments have been performed on 130 LiTS CT volumes only, having their corresponding manual segmentation data. The entire dataset is split into 70%, 15% and 15% for training, validation and testing sets respectively.

*B. Preprocessing*

The following preprocessing techniques are investigated and applied to enhance the raw CT images prior to feed it to U-Net architecture.

*1) Hounsfield Unit windowing*

It has been known that the HU value range for liver of a particular patient may varies, depending on the CT scanner [34]. In this regard, the researchers who study liver in medical images using deep learning, often limit the HU values range to focus on the target organ. In this study, one of the preprocessing techniques under investigation is HU value windowing. The HU value is limited to [-100, 400], which aims to differentiate liver from other adjacent organs.

*2) Contrast Limited Adaptive Histogram Equalization*

The quality of medical images obtained by different modalities can be improved by applying the adaptive histogram equalization technique, results in better segmentation at later stage. However, adaptive histogram equalization for CT image makes it too bright to see, therefore CLAHE is used for this study that overwhelmed the over-amplification problem [35]. The parameters chosen for CLAHE implementation are 4 as a clip limit and tile size of (8,8) for all experiments.

*3) Z-score normalization*

Z-score linearly transforms data or range of pixel intensity values to have zero mean and variance of 1. For this study, the z-score is obtained by subtracting the mean of pixels intensity values from original values, resulting in coinciding the mean on zero and finally the result is divided by standard deviation to equalize the spread. The data can be normalized using z-score as per giver formulae:

$$Z = \frac{x_i - \mu}{\sigma} \quad (1)$$

Where, Z is z-score normalized values. $x_i$ are the original values and $\mu$ is the mean of data values.

*4) Median filtering*

Median filter is a non-linear filter, in which the median of input values under a specific window size is computed as the output sample of the filter. The median filter involves horizontal window for one dimensional data value. However, for two-dimensional data values such as 2D image, spatial median filters are used with window or kernel size of 3x3. For this study, we have applied median filter to entire dataset in slice wise fashion to remove the pepper and salt noise. The kernel size chosen for median filtering is 3x3. Although, median filter removed noise from raw image, however, it produces quite good result by applying to HU windowed 2D image.

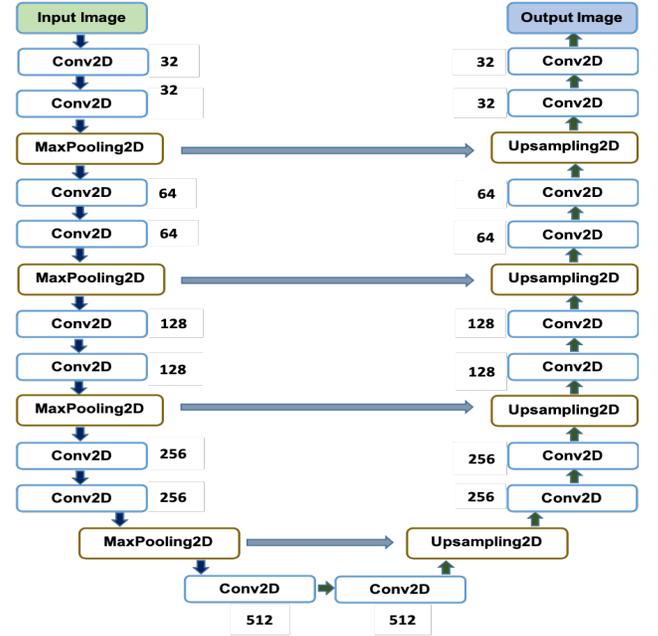

**Fig. 2.** A U-Net architecture. Each Conv2D operators implemented a kernel of size 3x3, ReLU as activation function, operation of 2x2 max pooling with stride 2 and same padding.

*5) Block-Matching and 3D filtering*

As a preprocessing stage, BM3D is another technique under investigation, applied to the entire dataset. It is a collaborative filtering process for image de-noising. In this method, the image blocks are processed in sliding manner, and search for blocks similar to a given block. The matched blocks are then stacked together, forming a 3D block. A 3D linear transformation is applied to all 3D blocks followed by Wiener filtering and Hard thresholding which results in denoising, and finally yields estimated 3D blocks by inverse 3D transform. This technique can be used to eliminate Gaussian noise and enhance signal-to-noise ratio in CT images.

**Table 1:** Summary of the hyperparameters used for training

| Hyperparameters | Values |
|---|---|
| Learning rate | 1e-4 |
| Epochs | 40 |
| Batch size | 32 |
| Optimizer | Adam |

## C. U-Net Architecture

U-Net architecture based on fully convolutional network (FCN), is widely used for biomedical image segmentation. Compared to other variants of FCN approaches, U-Net is symmetric and has multiple skip connections between down-sampling path and up-sampling path, which makes it enable to combine location information with contextual information. These multiple up-sampling layers make the U-Net a robust model for semantic segmentation. U-Net architecture as illustrated in Fig. 2, has 23 convolutional layers for both up-sampling and down-sampling paths in total. The implemented network architecture is slightly differing from the original U-Net model as proposed by Ronneberger et al. [5]. In the original U-Net model, 64 number of feature channels that begin the network, have been reduced to 32 channels as shown in Fig. 2. Also, the maximum number of feature channels at convolutional layers 9 and 10 are reduced to 512 from 1024 channels. Finally, a sigmoid function is used as activation function in the final 1x1 convolution, which ensures a range of [0,1] for mask pixels. A summary of hyperparameters with their respective values implemented in our algorithm is shown in Table 1.

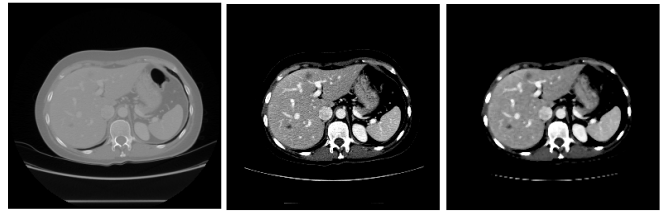

**Fig. 3.** Image preprocessing: Raw CT image (Left), HU windowed image (Middle) and median filtered image (Right).

**Table 2:** Result comparison of preprocessing techniques employed separately

| S. No | Preprocessing techniques | Results (Dice score) | | |
|---|---|---|---|---|
| | | Training % | Validation % | Testing% |
| 1 | HU windowing | 96.24 | 90.48 | 90.04 |
| 2 | CLAHE | 96.29 | 89.25 | 88.22 |
| 3 | z-score | 95.60 | 89.62 | 87.01 |
| 4 | Median filter | 95.24 | 89.96 | 88.82 |
| 5 | BM3D filtering | 94.45 | 89.12 | 88.34 |

**Table 3.** Testing result comparison of all sequences used in this work

| Sequence No. | HU windowing | CLAHE | BM3D filtering | Median filtering | z-score | Dice score (%) | | |
|---|---|---|---|---|---|---|---|---|
| | | | | | | Training | Validation | Testing |
| 1 | ✓ | ✓ | ✗ | ✗ | ✗ | 95.36 | 90.97 | 90.13 |
| 2 | ✓ | ✗ | ✗ | ✓ | ✗ | 96.46 | 90.39 | 90.24 |
| 3 | ✓ | ✓ | ✗ | ✓ | ✗ | 96.51 | 90.65 | 90.41 |
| 4 | ✓ | ✓ | ✗ | ✗ | ✓ | 94.76 | 90.99 | 89.10 |
| 5 | ✓ | ✗ | ✗ | ✗ | ✓ | 97.13 | 90.23 | 90.08 |
| 6 | ✗ | ✓ | ✗ | ✗ | ✓ | 96.95 | 88.90 | 88.75 |
| 7 | ✓ | ✗ | ✗ | ✓ | ✓ | **96.93** | **90.77** | **90.84** |
| 8 | ✗ | ✗ | ✗ | ✓ | ✓ | 96.80 | 90.66 | 89.24 |
| 9 | ✓ | ✗ | ✓ | ✗ | ✗ | 95.79 | 89.59 | 89.01 |
| 10 | ✓ | ✓ | ✓ | ✗ | ✗ | 95.82 | 90.31 | 88.92 |
| 11 | ✓ | ✓ | ✓ | ✗ | ✓ | 96.16 | 90.73 | 90.10 |
| 12 | ✓ | ✗ | ✓ | ✗ | ✓ | 95.93 | 88.84 | 89.19 |

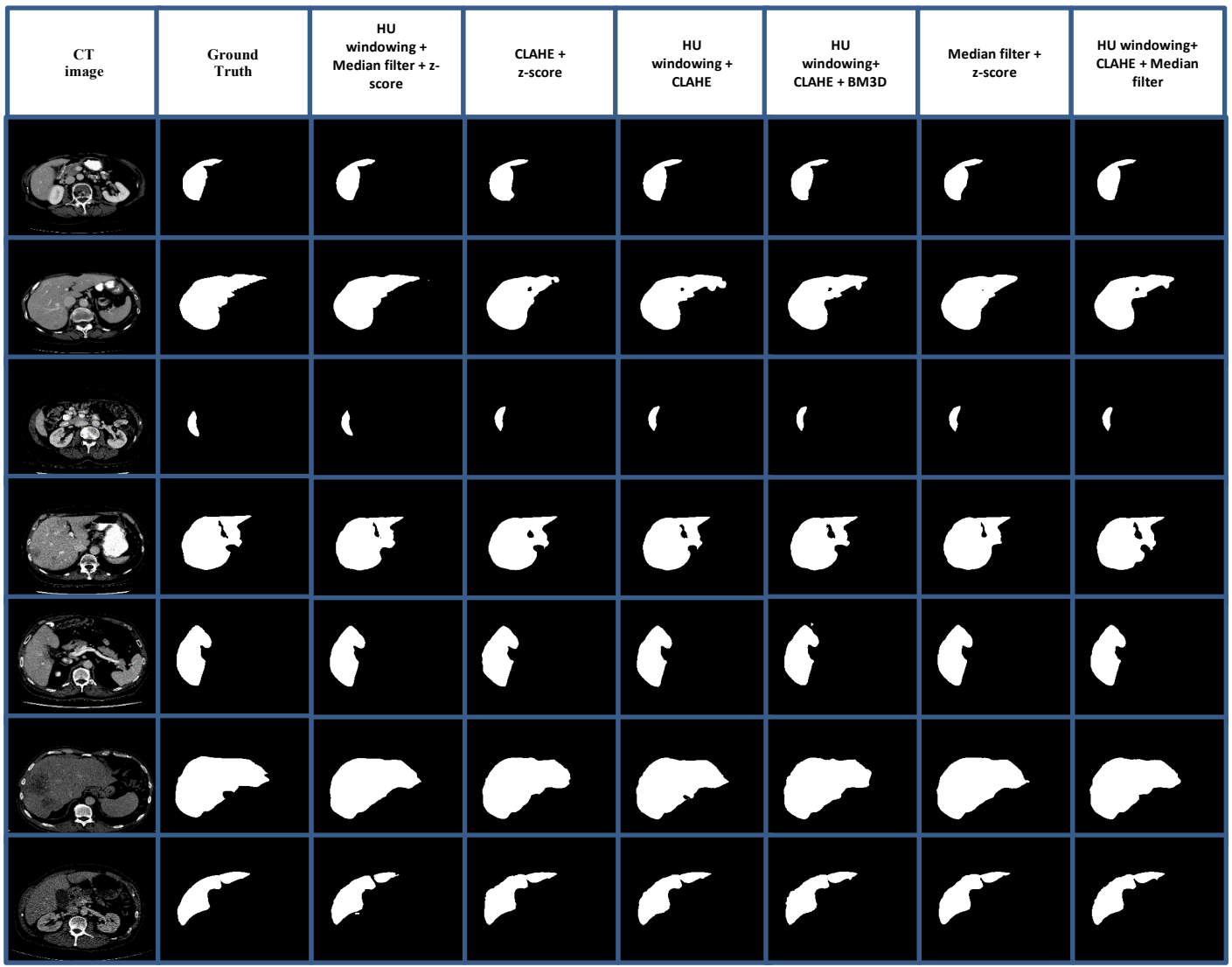

**Fig. 4**. Segmentation (prediction) results of proposed method along with other applied sequences.

## IV. RESULTS AND DISCUSSION

### A. Evaluation metric

To measure the model prediction accuracy, we used the most commonly used metrics known as Dice coefficient or Dice score and Dice loss as loss function. The Dice coefficient is considered as the matching level between predicted image and ground truth image. It ranges from 0 to 1, where 0 indicates no overlap or zero prediction while 1 indicates perfectly overlapping segmentation of liver. Dice coefficient and Dice loss can be calculated by the following equations:

$$Dice\ coefficient = \frac{2|X \cap Y|}{|X|+|Y|} \quad (2)$$

$$Dice\ Loss = 1 - \frac{2|X \cap Y|}{|X|+|Y|} \quad (3)$$

Where, X is the predicted image region and Y is the ground truth segmentation.

### B. Performance evaluation

In this work, different preprocessing techniques are compared for automated liver segmentation. Experiments are performed using preprocessing techniques including HU windowing, CLAHE, z-score, median and BM3D filtering. In Table 2, the performance results of each preprocessing technique are listed in terms of training, validation and testing Dice coefficients. Here, only one technique is employed at a time in order to investigate the effects of preprocessing techniques separately. From Table 2, it can be inferred that HU windowing performed the best as a preprocessing technique with 96.24%, 90.48%, and 90.04% training, validation and testing dice coefficient respectively. Hence, HU windowing is an important preprocessing method which needs to be applied before medical image segmentation task.

In the second stage, combinations of different preprocessing techniques are employed to evaluate its performance. In this work, twelve sequences are chosen shown in Table 3, to study their impacts on liver segmentation. In the table, (✔) means that corresponding technique is applied while (✘) shows that the technique is not applied to the given sequence. The order of employed preprocessing techniques are from left to right, means that in any sequence the first employed technique is HU windowing, followed by other techniques in the given order. Comparing all the sequences with each other i.e., sequence-1 to sequence-12, it can be concluded that the best result shown by sequence-7, which is a combination of HU windowing followed by median filtering and z-score orderly. The image before and after employed sequence-7 is shown in Fig. 3. Similarly, followed sequence-7, the second sequence shown best result is sequence-3 i.e., HU windowing followed by CLAHE and median filtering.

## C. Prediction results

Different combinations of preprocessing techniques are investigated and compared their impact on liver segmentation in CT images. The prediction results of first six sequences are depicted in Fig. 4, in order to validate the proposed method (sequence-7). For predication purposes, a few images from LiTS dataset are randomly selected along with corresponding ground truth images. The prediction results confirmed that the proposed sequence performed the best among the other sequences.

## V. CONCLUSION AND FUTURE WORK

Preprocessing is an essential procedure to enhance the quality of input training data for training deep learning models that segment liver from various medical images. In our work, the impact of different preprocessing techniques is analyzed by employing various preprocessing techniques and proposed the optimal preprocessing technique that performed comparatively well in network training for automated liver segmentation. The proposed framework performed well when the input data is preprocessed with combination of three techniques i.e., HU windowing followed by median filtering and z-score normalization and gave 90.84% Dice score for test dataset. In future, this work can be extended further for segmentation of liver lesions and other body organs in medical images.

## ACKNOWLEDGEMENT

We gratefully acknowledge the support of Artificial Intelligence in Healthcare, Intelligent Information Processing Lab, National Center of Artificial Intelligence, UET Peshawar for the necessary research infrastructure and support. We are also grateful of NVIDIA Corporation for supporting this research by providing Titan X Pascal high computing facility